\documentclass[runningheads]{llncs}

\usepackage{soulutf8}
\usepackage{url}
\usepackage{amsmath,amssymb,amsfonts}
\usepackage{algorithmic}
\usepackage{graphicx}
\usepackage{textcomp}
\usepackage{xcolor}
\usepackage{tabto}
\usepackage{mdframed}
\usepackage{booktabs}
\usepackage{multirow}
\usepackage{makecell}
\usepackage{subcaption}
\usepackage[hidelinks]{hyperref}
\usepackage[numbers,sort]{natbib}

\begin{document}



\title{A Text Mining and Classification Approach for Analyzing Architecture Decision Records}
\titlerunning{A Text Mining and Classification Approach for Analyzing ADRs}







\author{Nicolas Miccio\inst{1} \and
Antonela Tommasel\inst{1,2} \and
J. Andres Diaz-Pace\inst{1}}

\authorrunning{N. Miccio et al.}

\institute{
ISISTAN Research Institute, CONICET-UNCPBA, Tandil, Argentina,  
\and
Johannes Kepler University, Linz, Austria\\
\email{\{nicolas.miccio,antonela.tommasel,andres.diazpace\}@isistan.unicen.edu.ar}
}

\maketitle

\begin{abstract}
Architectural decision records (ADRs) have become a popular lightweight mechanism for documenting architectural knowledge in software projects. However, there is limited empirical evidence on the kinds of architectural concerns captured in ADRs and how well their contents align with established architectural knowledge concepts and documentation practices.
In this paper, we propose an automated text-mining and classification approach for analyzing ADRs at scale. We apply this approach to a dataset of ADRs extracted from $\approx550$ open-source repositories, combining topic modeling, LLM-based classification, and template compliance checks. Our analysis examines decision taxonomies and quality attributes, and the degree to which ADRs adhere to the MADR template.
Our findings show that ADRs frequently capture existence, technology, and process-related decisions, while alternatives, decisions drivers, and some quality concerns remain under-documented. We also observe recurring mismatches between ADR contents and template sections. 
These insights into current documentation practices provide architects with valuable information to reflect on how ADRs are and should be used to effectively deal with architectural knowledge. Furthermore, our  automated approach is adaptable to other architectural tasks.
\end{abstract}


\section{Introduction}\vspace{-0.25cm}

Capturing design decisions is a valuable practice in software projects as it helps preserve the rationale, constraints, and trade-offs that shape system evolution. In the context of software architecture, this knowledge is often referred to \textit{Architecture knowledge} (AK) \cite{Capilla2016}. A popular lightweight approach for AK management is the use of \textit{architecture decision records} (ADRs) \cite{Keeling2022, ahmeti2024architecture}, which provide a simple textual mechanism for documenting key design decisions and their rationale, often through templates with required and optional sections, such as the \textit{Markdown Architectural Decision Record} (MADR) format\footnote{\url{https://adr.github.io/madr/}}.

Despite their growing adoption, there is still limited empirical evidence on how ADRs are actually used in practice. Little is known about the concerns practitioners choose to document, how these concerns align with established architectural decisions taxonomies \cite{kruchten2004ontology, ZIMMERMANN20091249}, or whether ADRs adequately capture aspects like rationale, alternatives and quality attributes. Existing studies have mainly focused on ADR adoption and usage patterns \cite{buchgeher2023using}, leaving ADR contents largely unexplored. This gap matters as incomplete or low-quality ADRs hinder decision traceability, system evolution, and the long-term AK preservation.

To address this gap, we conduct an empirical study of ADR contents using a dataset of open-source projects \cite{buchgeher2023using}. We examined $\approx550$ projects with a total of $\approx4300$ ADRs from three complementary perspectives: (i) the concerns discussed in ADRs, (ii) their alignment with decision taxonomies and quality attributes, and (iii) their structural consistency with key decisions of the MADR template. For this purpose, we define a text-mining and classification pipeline combining topic modeling and LLM-based analysis.
Our findings show that ADRs frequently document existence, technology, and process-related decisions, while design alternatives, decision drivers, and some quality concerns remain under-documented. We also identify mismatches between ADR contents and the intended purpose of sections. These results contribute empirical insights into current ADR documentation practices and suggest opportunities to improve ADR authoring via tool support, and demonstrate how LLM-based analysis pipelines can support large-scale architectural knowledge studies.

\vspace{-0.55cm}\section{Background and Related Work}\vspace{-0.1cm}
\subsubsection*{Architecture Decision Records.}\vspace{-0.25cm}

\begin{figure}[t]
\centering
\includegraphics[width=\linewidth]{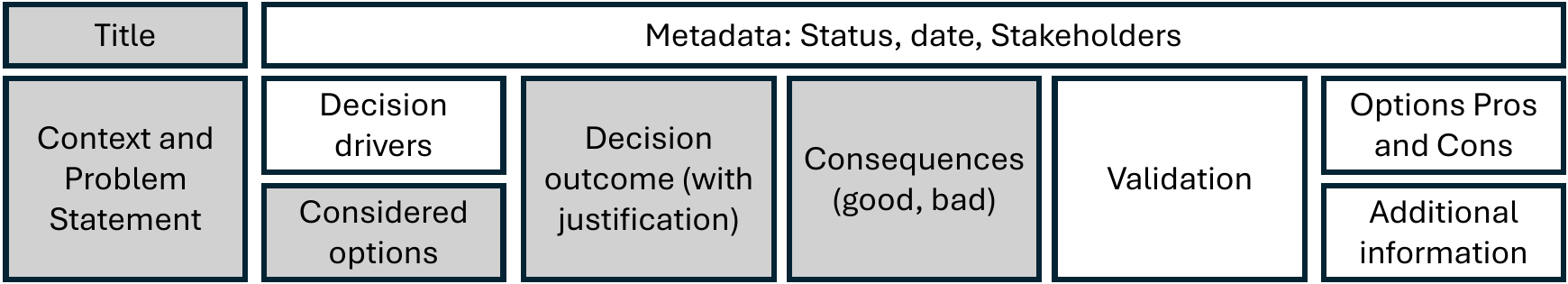}\vspace{-0.2cm}
\caption{Representation of a typical ADR template (MADR)}\label{fig:adr-template}\vspace{-0.65cm} 

\end{figure}

Architectural decisions are design choices that address a functional or quality-attribute concern and have a significant impact on the system.
Such significance typically arises from the decision’s long-term consequences or the high cost of modifying it in later developement stages. An ADR documents a key decision together with its rationale, enabling teams to capture and communicate AK over time. ADRs were popularized by M. Nygard and others \cite{Keeling2022} as lightweight, text-based artifacts stored alongside source code. Their simplicity 
has contributed to their growing use in software projects \cite{Capilla2016, Zdun2013}.
Among existing templates, MADR (exemplified in Fig. \ref{fig:adr-template}\footnote{Adapted from: \scriptsize{\url{https://ozimmer.ch/practices/2022/11/22/MADRTemplatePrimer.html}}}) has become a widely used format for ADRs. In addition to core sections such as \textit{Context and Problem Statement}, \textit{Decision Outcome}, and \textit{Consequences}, MADR also includes \textit{Decision Drivers}, \textit{Considered Options} and \textit{Options Pros and Cons}, aiming to better support rationale capture, traceability and comparison of alternatives.

\vspace{-0.25cm}\subsubsection*{Previous Works on ADR Usage.}\vspace{-0.25cm} 

Despite their appeal, ADRs are not yet consistently used in practice. Earlier works identified barriers such as time constraints, weak process integration and limited tool support for documenting decision rationale \cite{falessi2008value}. More recently, Buchgeher et al. \cite{buchgeher2023using} mined ADR usage across GitHub repositories and found that, while ADR adoption is increasing in open-source projects, many repositories contain only a few ADRs and use them in inconsistent or ad-hoc ways. Their study focused primarily on adoption patterns, template usage, and evolution over time, leaving the actual contents of ADRs largely unexplored. 
Other studies have reported organizational inconsistencies in ADR authoring, uncertainty about what qualifies as an architecturally significant decision, and difficulties in maintaining ADRs over time \cite{ahmeti2024architecture}. In parallel, recent work has explored the use of LLMs to generate or improve ADRs \cite{dhar2025draft, diaz2024helping}. 
While these approaches show promise for supporting ADR authoring, they do not provide empirical evidence about the concerns 
that ADRs capture. 

Overall, the studies so far have advanced the understanding of how ADRs are used and how they can be generated, but an important gap remains: \textit{what do ADRs talk about?} 
This paper takes a step towards addressing this gap by analyzing the thematic and structural content of ADRs at scale, considering the types of decisions, concerns, and quality attributes.

\begin{table}[t]
    \centering
        \caption{Kruchten's \cite{kruchten2004ontology} architectural decision ontology}
    \small
       \begin{tabular}{p{0.17\columnwidth}|p{0.80\columnwidth}}
        \textit{Existence} (ontocrisis) & These decisions declare that an element or artifact will exist in the design or implementation. Includes structural decisions (e.g., layers, components) and behavioral decisions (e.g., connectors, interactions). Structural decisions lead to the creation of subsystems, layers, partitions, or components. 
        Behavioral decisions are more related to how the elements interact together to provide functionality or to satisfy some non-functional requirement (quality attribute), or connectors. \\
        \hline
        \textit{Ban/Non-existence} (anticrisis) &
        These decisions 
        declare that an element will not exist in the design or implementation and often used to rule out alternatives.\\
        \hline
        \textit{Property} (diacrisis)  & These decisions state a general, enduring quality or constraint of the system. These are often cross-cutting concerns or design rules (positive) or constraints (negative). \\
        \hline
        \textit{Executive} (pericrisis) & These decisions refer to decisions that do not relate directly to the design elements or their qualities but are driven more by the business environment (financial), and affect the development process (methodological), the people (education and training), the organization, and to a large extend the choices of technologies and tools.\\
        
    \end{tabular}
    \label{tab:kruchten-ontology}\vspace{-0.65cm}
 
\end{table}

\begin{table}
    \centering
\caption{Zimmermman classification of decisions based on their main concern }
    \hspace{-0.3cm}
    \small
    \begin{tabular}{p{0.19\columnwidth}|p{0.75\columnwidth}}
        \textit{Design}  & They concern the logical organization, structure, and decomposition of the system. Related to patterns, components, layering, interfaces, data modeling.\\
        \hline
        \textit{Technology} & They concern the selection of technologies, platforms, frameworks, libraries, or standards.\\
        \hline
        \textit{Infrastructure} & They involve the deployment environment, hosting, runtime platforms, networking, and hardware concerns.\\
        \hline
        \textit{Organizational Process} & They concern team structures, roles, responsibilities, workflows, and processes that affect architecture.\\
        \hline
        \textit{Constraints} & They refer to mandatory conditions from the business, regulations, or existing systems that limit architectural choices.\\
       \hline
       \textit{Quality-Attribute} & They explicitly target system qualities like performance, security, availability, etc. These are decisions made primarily to meet a non-functional requirement.\\
        \hline
        \textit{Crosscutting Concern}s & They relate to decisions that impact multiple parts of the system simultaneously, often aspects like logging, monitoring, security mechanisms.\\
        \hline
        \textit{Implementation} & They affect internal code structure, patterns at the class or method level, or maintainability mechanisms, but are not architectural in scope.\\
    \end{tabular} 
    \label{tab:architectural-decision-focus}\vspace{-0.75cm}
  
\end{table}

\vspace{-0.3cm}\subsubsection*{Types of Architecture Decisions.}\label{sec:decision-taxonomy}\vspace{-0.25cm}

To characterize ADR contents, we rely on established taxonomies from the AK literature. First, we use Kruchten’s ontology of architectural decisions \cite{kruchten2004ontology}, which distinguishes among \textit{existence}, \textit{property}, and \textit{executive} decisions (Table~\ref{tab:kruchten-ontology}). Prior evidence suggests that existence decisions are the most frequently documented in practice ($\approx 65\%$), while executive and property decisions are less common ($27\%$ and $8\%$, respectively) \cite{miesbauer2013classification}. Notably, non-existence decisions were not identified, despite their importance. As emphasized in \cite{kruchten2004ontology}, such decisions are critical to document because they are not evident in the system structure or implementation and can easily be overlooked. 

Second, we use Zimmermann’s classification of architectural decisions by \textit{main concern or focus} 
\cite{ZIMMERMANN20091249}, which includes categories such as design, technology, infrastructure, implementation, organizational/process, and quality-attribute-related decisions (Table~\ref{tab:architectural-decision-focus}). This taxonomy helps characterize what ADRs are primarily about.
Lastly, as quality attributes are central to architectural reasoning, we also analyze whether ADRs address concerns such as performance, reliability, security, maintainability, scalability, usability, portability, and testability. These attributes are grounded in the \textit{ISO/IEC 25010 Standard for System and Software Quality Models}\footnote{\url{https://www.iso25000.com/index.php/en/iso-25000-standards/iso-25010}} and complemented with 
\cite{sap:bassetal2012}. Together, these three perspectives provide a structured basis for assessing ADR contents.

\begin{figure}[t]
\centering
\includegraphics[width=0.92\linewidth]{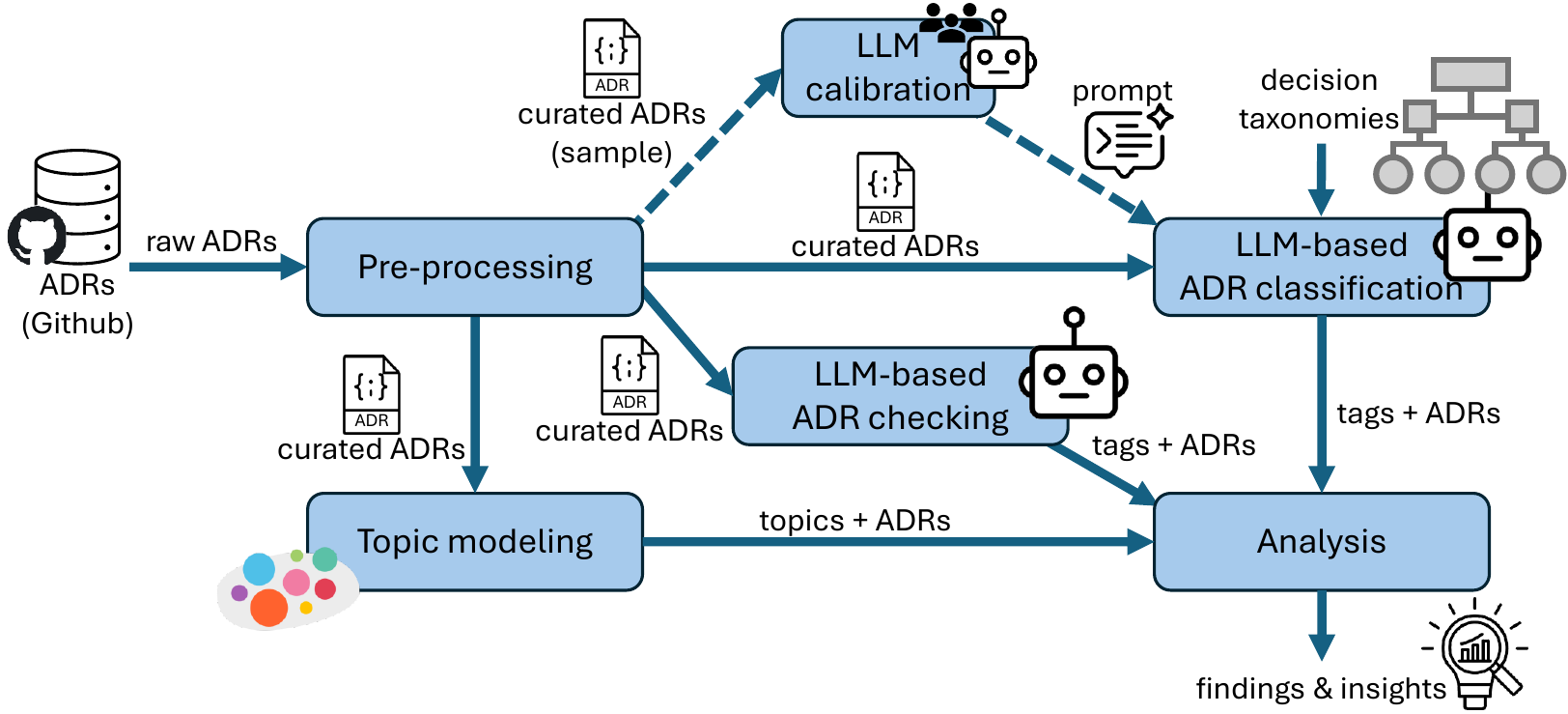}\vspace{-0.4cm}

\caption{ADR mining and processing pipeline. 
}\label{fig:approach}\vspace{-0.7cm}
\end{figure}

\vspace{-0.35cm}\section{Study Design}\vspace{-0.35cm}

The large number of ADRs in our dataset makes manual examination and annotation infeasible. At the same time, relying solely on statistical text-mining techniques is often not sufficient to capture the architectural intent and decision semantics embedded in ADRs. To balance scalability and interpretability, we therefore designed an automated analysis pipeline that combines topic modeling with LLM–based classification and checking. This hybrid approach enable us to uncover recurring themes in ADRs while mapping their contents to established architectural decision taxonomies and template structures.

The pipeline consists of six components, as outlined in Figure \ref{fig:approach}. 
1) \ul{\textit{Pre-processing.}} It involves the collection of ADRs from projects in open-source repositories, an explorative data analysis of the ADRs (documents), and the application of exclusion and inclusion criteria for our study. The outcome of this stage is a curated dataset. 
2) \ul{\textit{Topic modeling.}} It performs an unsupervised analysis of the dataset to identify latent topics in the documents. We apply embeddings on the textual contents of the ADR documents and group them using clustering techniques. 
3) \ul{\textit{LLM prompting and calibration.}} It generates a small ground-truth dataset used to assess classification quality and calibrate the LLM-based labeling process.
4) \ul{\textit{Classification based on decision taxonomies.}} It maps the contents of every document to predefined decision categories according to the three taxonomies mentioned in Section 2 and apply text classification aided by an LLM. 
5) \ul{\textit{Checks using MADR template.}} It scans the documents and identifies correspondences with the MADR sections, determining whether each section is present and fulfills its intended purpose. This evaluation goes beyond standard text classification and leverages an LLM-based approach. 
6) \ul{\textit{Analysis and insights.}} Based on the outputs of the previous components, both quantitative and qualitative analyses are performed. These analyses were driven by our research questions.\vspace{0.1cm}

We tackle the following research questions: 
\textbullet~\textbf{RQ\#1.} \textit{What types of 
concerns (as topics) are captured in 
ADRs?} We explore concerns in a broad sense, without ties to existing taxonomies or architectural concepts. 
\textbullet~\textbf{RQ\#2.} \textit{How well does the captured content align with the taxonomies from the literature?} We analyze the proportion of ADRs that reflect Kruchten’s decision types, examine their primary concerns using Zimmermann's framework, and identify which quality attributes appear most or least frequently. 
\textbullet~\textbf{RQ\#3.} \textit{What are common mismatches or inconsistent practices in the usage of the ADR structure (i.e., MADR sections)?} 
We assess whether section titles accurately reflect their corresponding content, and also identify if these sections are omitted or misused.

In our study, we followed the guidelines for mining software repositories from \cite{vidoni2022systematic}, which comprise: definition of sources, search for repositories, data extraction, and information generation.
All LLM-related experiments were conducted with \texttt{GPT-4.1-mini} model\footnote{This model was selected because it provided a practical trade-off between reasoning capability, structured-output reliability, latency, and cost for large-scale prompt-based ADR analysis. Since our goal was not to benchmark LLMs, but to study the feasibility of automated ADR classification, we prioritized scalability and stable instruction-following behavior over model comparison.}. The LLM was accessed via the \texttt{LangChain} framework\footnote{\url{https://python.langchain.com}} with a temperature set to $0$ to minimize randomness in responses\footnote{\url{https://platform.openai.com/docs/api-reference/chat}}. Each ADR was processed independently 
without conversational memory. 

\vspace{-0.5cm}\subsection{Data Collection}\vspace{-0.25cm}

Our dataset of ADRs from open-source GitHub repositories was originally compiled by Buchgeher et al. \cite{buchgeher2023using} and covers projects in the 2013-2020 period. The dataset, provided in JSON format, includes metadata such as repository URLs and ADR file locations. Since our analysis does not focus on temporal evolution, we retrieved the latest available version of each ADR. If there were no ADRs in the most recent commit, the last commit referenced in the original dataset was retrieved. In some cases, repository names had changed and required manual correction. The ADRs were originally identified through repository mining heuristics based on common ADR conventions, including dedicated ADR directories and filename or heading patterns associated with architectural decision records. We further refined the dataset through preprocessing and filtering steps aimed at removing duplicated, empty, or template-only documents.

We performed a cleaning step to ensure data quality. ADRs were discarded if they were duplicated, not written in English, or if they were only 
templates for creating real ADRs. Additionally, we filtered out projects containing fewer than five ADRs or whose ADRs were shorter than $500$ characters on average, under the assumption that they would provide insufficient content for our analysis. From the original $867$ projects we retrieved $6061$ ADRs. After filtering, our final dataset consists of $4316$ ADRs from $547$ (unique) projects.

\vspace{-0.5cm}\subsection{Preprocessing}\vspace{-0.3cm}

To extract relevant information from the ADRs, we developed a custom Markdown parser. 
Despite Markdown's standardized syntax, 
the variability in how organizations structure ADRs posed 
parsing challenges. While some ADRs followed well-defined templates like MADR, others deviated considerably. 
We adap\-ted the parser to identify such cases to avoid missing key content. 
Heading levels were also applied inconsistently, which complicated structural interpretation of sections. 
Extracting the main \textit{Decision Outcome} section proved particularly challenging, as it could appear at various levels of the hierarchy, be embedded within broader sections, or be accompanied by subsections like \textit{Pros}, \textit{Cons}.

\vspace{-0.5cm}\subsection{Topic Modeling}\vspace{-0.25cm}

\texttt{BERTopic} \cite{grootendorst2022bertopic}\footnote{\url{https://maartengr.github.io/BERTopic/index.html}} is a topic modeling approach that leverages transformers, clustering and TF-IDF techniques to generate coherent topic representations of documents. In our context, a document is equivalent to an ADR; thus, we can extract the main topics covered by the ADRs. 
In particular, \texttt{BERTopic} first creates document embeddings to obtain document-level information, and then reduces the dimensionality of the embeddings before creating semantically similar clusters, each one representing a distinct topic. A class-based variation of TF-IDF is used to extract the topic representation from each cluster. Basically, all documents within a cluster are treated as a single document or class (by concatenating their contents). A TF-IDF strategy is adjusted to account for this representation. This procedure, called \textit{c-TF-IDF}, permits to capture the importance of words in clusters rather than individual documents. 

Traditional topic modeling techniques such as Latent Dirichlet Allocation (LDA) were considered, but they rely on bag-of-words assumptions that often produce sparse and less coherent topics for short, domain-specific texts like ADRs. \texttt{BERTopic} was preferred because it leverages contextual embeddings from transformer models to capture semantic similarities across heterogeneous vocabulary, yielding more meaningful and fine-grained clusters for our artifacts. 

To assess topic coherence and architectural relevance, two authors manually inspected a sample of $5$ ADRs associated with each topic, together with their keyword representations. In this qualitative validation, we analyzed that the sampled documents per topic cluster were consistent and that the topic labels were representative of the cluster. While this step was not intended as a full qualitative coding exercise, it provided confidence that the identified topics were meaningful from an architectural perspective.

\vspace{-0.5cm}\subsection{Prompting and LLM Calibration}\vspace{-0.2cm} 

LLMs enable classification through instructions rather than task-specific training \cite{brown2020language, yu2022generate}. This makes LLMs suitable for our context, where little (or no) labeled data is available and ADR content varies across repositories and projects. Given the unstructured format of ADRs and the scale of our dataset, manual annotation of all ADRs would have been very time-consuming and impractical. For these reasons, we designed an LLM-based classification schema with prompt instructions for our three taxonomies, and applied it across the entire ADR corpus.

\vspace{-0.35cm}\paragraph{\underline{Ground-truth construction.}}
To support the evaluation of the automated classification, a subset of ADRs was considered. We selected a random sample of $200$ ADRs from different projects and manually annotated them according to the taxonomies (Kruchten's, Zimmermann's and quality attributes). Two of the authors acted as reviewers and independently labeled $125$ ADRs each. To assess annotation consistency, $50$ overlapped ADRs were evaluated by both authors. 
Disagreements and misclassifications were subsequently discussed until consensus was reached. This annotated subset constitutes a (small) \textit{ground-truth }dataset for the study, although not necessarily covering all the ADR cases. 

\vspace{-0.35cm}\paragraph{\underline{Embedding-based baseline.}}
Using the ground-truth dataset, we first investigated whe\-ther a traditional supervised classification schema would be sufficient for the task. For each taxonomy, we trained a multi-class classifier based on gradient boosting (\textit{XGBoost}) using document embeddings (\textit{allMiniLM-L6-v2}) extracted from ADR texts. 
However, this schema yielded limited performance. The classifiers exhibited strong sensitivity to class imbalance, favoring majority classes (e.g., existence decisions in Kruchten's taxonomy; maintainability and security for quality attributes; and design and technology in Zimmermann’s framework). Moreover, confusion-matrix analysis revealed poor separability among classes, indicating that embeddings alone are insufficient to capture the semantic intent and pragmatic function required for taxonomy-based ADR classification. These limitations motivated the adoption of an LLM-based classification schema. 

\vspace{-0.35cm}\paragraph{\underline{LLM prompt refinement.}}
For the ADRs in the ground-truth dataset, we compared the labels assigned by human experts with those inferred by the LLM. 
In addition to the system prompt, all taxonomies shared a common prompt structure, into which taxonomy-specific definitions, rules, and examples were injected. On this basis, we evaluated three prompting strategies\footnote{Examples of prompt evolution are available in the companion repository: \url{https://github.com/tommantonela/ADRminer}}, following an iterative prompt-refinement process\footnote{\url{https://llm-guidelines.org}} for improving classification reliability.

The first strategy corresponded to a \textit{zero-shot} configuration, in which the LLM was asked to assign a primary category to an ADR based solely on the taxonomy description and instructions, without examples or additional architectural guidelines. Although the task was formulated as single-label classification, the prompt allowed the model to return alternative categories together with qualitative confidence levels in a structured JSON response. An analysis of discrepancies between human annotations and zero-shot LLM outputs revealed several weaknesses in the initial prompt formulation. Common confusion sources included the distinction between executive and existence decisions in Kruchten’s taxonomy; the dominance of maintainability over related quality attributes such as portability or compatibility; and subtle boundaries between design, technology, and implementation concerns in Zimmermann’s framework. Organizational and process-related decisions were also occasionally misclassified as product-level concerns, further motivating the need for more explicit prompt guidance. 

For the second strategy, as zero-shot exhibited limited agreement with the ground truth, we extended the prompts with \textit{static few-shot} examples, providing representative ADR excerpts per category. We introduced also chain-of-thought-style reasoning instructions and disambiguation rules for frequently confused categories. 
As a third strategy, we employed \textit{dynamic few-shot} prompting, in which a small set of examples were automatically selected from the ground-truth (based on semantic similarity to the input ADR) and injected into the prompt. For this strategy, we also used the chain-of-though prompting mentioned above.

\vspace{-0.4cm}\subsection{LLM-based ADR Classification} \label{sec:classification}\vspace{-0.3cm}

Each ADR, after preprocessing, was provided to the LLM in its entirety. 
The model was instructed to return structured outputs containing: (i) the primary selected category, (ii) a brief explanation justifying the classification, and (iii) a confidence score (in the $0-1$ range) for each alternative category. 
To increase consistency and limit hallucinations, the LLM was instructed to rely on information explicitly stated in the ADR and to avoid inferring unstated intent.

Using the prompting configurations defined in the previous section, we applied the LLM-based classifier to the ADRs in the ground-truth dataset. Across the evaluated taxonomies, we observed agreement rates ranging from $46\%$ to $66\%$ for Kruchten’s taxonomy (4 classes), Zimmermann’s taxonomy (8 classes), and the quality-attribute classification (10 classes), across the evaluated prompts. As expected, classification quality varied across taxonomies and was influenced by the adopted prompting strategy, with higher agreement generally achieved when moving from zero-shot to few-shot configurations.

Overall, fewer discrepancies were observed for Kruchten’s taxonomy, more for Zimmermann’s categories, and an intermediate behavior for quality attributes. These differences can be attributed to both the stochastic LLM nature and the inherent subjectivity of certain taxonomies. For example, an ADR may address multiple quality attributes simultaneously or combine decisions involving technology and implementation details, which complicates classification. 
Based on a comparative evaluation of the prompting strategies, we selected the best-performing configuration for each taxonomy. In particular, the static few-shots proved effective for quality-attribute classification, while dynamic few-shots yielded the best performance for Kruchten’s and Zimmermann’s taxonomies, albeit at the cost of increased latency and token consumption. 

\vspace{-0.4cm}\subsection{LLM-based ADR Checking}\vspace{-0.25cm}

To analyze ADR structural consistency, we focused on five key MADR sections: \textit{Context and Problem Statement}, \textit{Decision Outcome}, consequences, \textit{Decision Drivers}, and \textit{Considered Options}. For each section, the following criteria were defined: \textbullet~\textit{Presence}. Whether a given section (i.e., its heading) is explicitly present in the ADR. 
\textbullet~\textit{Alternative title}. When a section did not use the expected MADR heading, we analyzed whether semantically equivalent content appeared under another heading in the ADR. In such cases, the detected alternative heading was recorded.
\textbullet~\textit{Content quality}. For each section (or its detected alternatives), we assessed whether the text had meaningful, project-specific content rather than vague or generic statements. 
\textbullet~\textit{Purpose consistency}. We examined whether the content fulfilled the section's intended purpose, as well as with the roles of other sections. We used a three-level scale for this criterion. A \textit{Yes} value indicated that the content was clear and well-scoped, fulfilling the intended section role. A \textit{Partial} value was assigned when some degree of overlap with another section was observed. Finally, a \textit{No} value denoted that the content either belonged to another section or failed to fulfill the section's intended purpose.\vspace{0.1cm}

These checks were implemented using an \textit{LLM-as-a-judge} strategy \cite{zheng:2024}. For ADRs in which the target section heading was present, the LLM was asked to evaluate content quality and purpose consistency, and provide a justification for the responses. When the heading was absent, the LLM was prompted to search for suitable content within the ADR text. Furthermore, the LLM received a description of the section’s intended purpose. Each section was evaluated independently to reduce cross-section interference. Throughout the analysis, the LLM was asked to strictly adhere to our evaluation guidelines, namely: use all available information from the ADR text, assume minimal external context, favor clarity over guesswork, and be conservative in section assessments. Sections that were duplicated or overlapped with others were penalized accordingly, and placeholders were marked as misuse unless appropriately replaced with project-specific information. To assess the reliability of the LLM-as-a-judge evaluations, two of the authors independently reviewed a random sample of $100$ ADR assessments. The review covered section detection, content quality, and purpose-consistency judgments according to the predefined evaluation criteria. The reviewers compared their assessments, discussed disagreements, and reached consensus on ambiguous cases. This process also revealed minor weaknesses in the initial checking prompts, which were subsequently refined.

\vspace{-0.55cm}\section{Analysis Results} \label{analysis}\vspace{-0.1cm}

\vspace{-0.25cm}\subsection{RQ\#1 - Types of Concerns as Topics}\vspace{-0.25cm}

\begin{figure*}[t]
\centering
\includegraphics[width=0.99\linewidth]{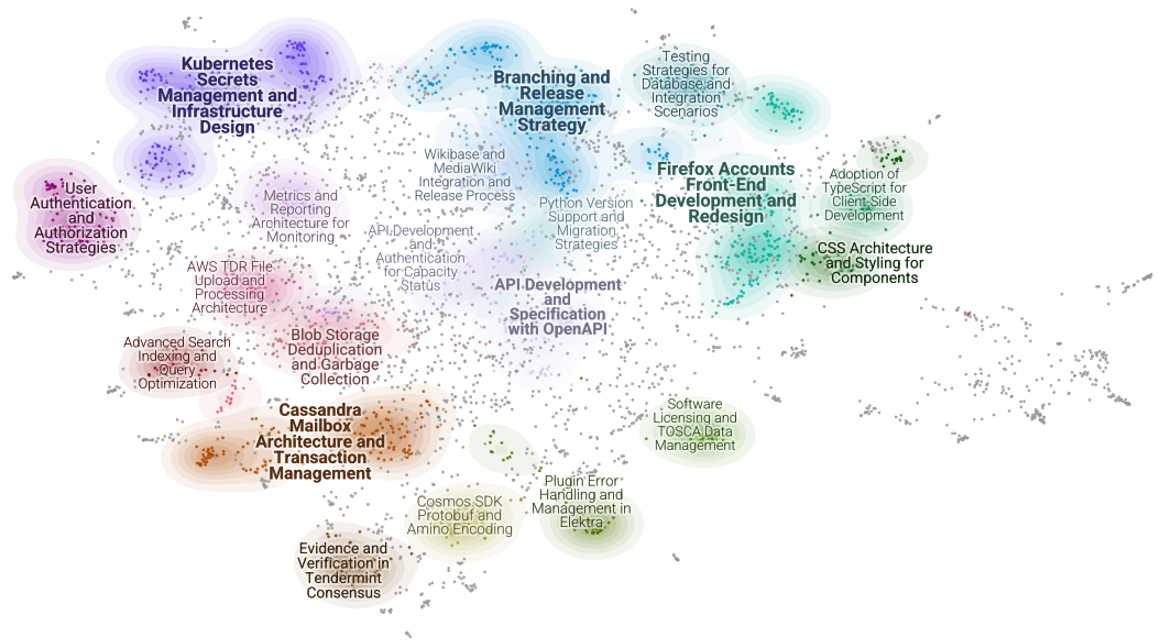}\vspace{-0.4cm}
\caption{Top-20 topics detected in the ADR contents (topic font size is proportional to the topic prevalence in the corpus)}\label{fig:bertopic}\vspace{-0.75cm}
\end{figure*}

Figure \ref{fig:bertopic} shows the main topics identified in the ADR corpus using \texttt{BERTopic}. The model produced $67$ topics after iterative topic reduction, yielding a reasonable balance between coherence ($0.38$) and diversity ($0.96$)\footnote{Perplexity was not computed because it is designed for probabilistic topic models such as LDA and is not directly applicable to embedding-based approaches like \texttt{BERTopic}. Instead, we evaluated topic quality using coherence and diversity metrics.}. To reduce the initial number of topics, \texttt{BERTopic} applies HDBSCAN over the c-TF-IDF topic representations and automatically merges semantically similar topics. The topics are visualized as clusters of related ADRs. Although a document (i.e., an ADR) can be affected by several topics, 
only the most prevalent topic is shown. Those ADRs with unclear predominant topics are marked as noise (gray points), often lying in the cluster boundaries. Overall, the topics were semantically distinct while still capturing recurring patterns across ADRs. 

Based on the keywords (topic representation) returned by \texttt{BERTo\-pic} for each topic, we assigned descriptive labels to the detected topics. Prominent topics included \textit{Authentication and Authorization}, \textit{Deployment}, \textit{Branches and Releases}, \textit{API Development}, \textit{NoSQL Databases}, \textit{Blockchain}, and \textit{Frontend}. A list of topic labels and representations is provided in Table~\ref{tab:topic_table}. 

We grouped the topics into categories by leveraging a hierarchical topic organization feature provided by \texttt{BERTopic}. The identified topics reveal that ADRs cover a mixture of architectural concerns, technology choices, and development-related activities. Recurring themes included \textit{databases}, \textit{security}, \textit{cloud infrastructure}, \textit{deployment}, and \textit{software development tasks} such as APIs, branching, telemetry, and programming languages. While some of these concerns map directly to established AK categories (e.g., deployment to infrastructure, authentication to security, frontend to design/usability), others reflect broader engineering practices that are not traditionally emphasized in AK models. This suggests that ADRs are often used not only to capture architectural structures, but also to document supporting technologies and crosscutting development concerns.

\begin{table*}[t]
\vspace{0.5cm}
\caption{Summary of the top-20 topics extracted from the ADRs. For better readability, topic representations were linked to descriptions, and descriptions were grouped into categories. 
}
\hspace{-0.25cm}~\includegraphics[width=1.01\columnwidth]{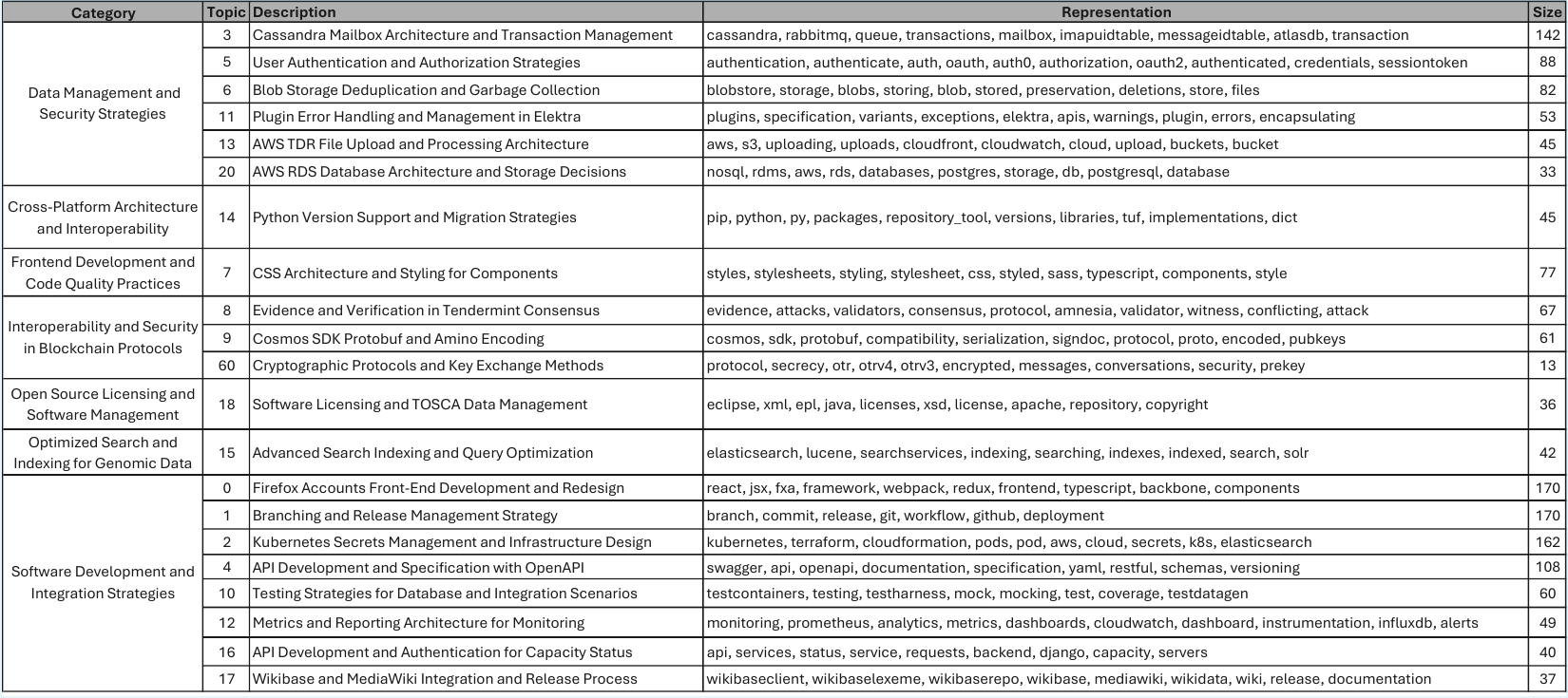}

\vspace{0.2cm}
\label{tab:topic_table}\vspace{-0.6cm}
\end{table*}

\vspace{-0.45cm}\subsection{RQ\#2 – Relations with Decision Taxonomies}\vspace{-0.25cm}

\begin{figure*}[t]
    \centering
    
    \begin{subfigure}[t]{0.32\textwidth}
        \centering
        \includegraphics[width=\linewidth]{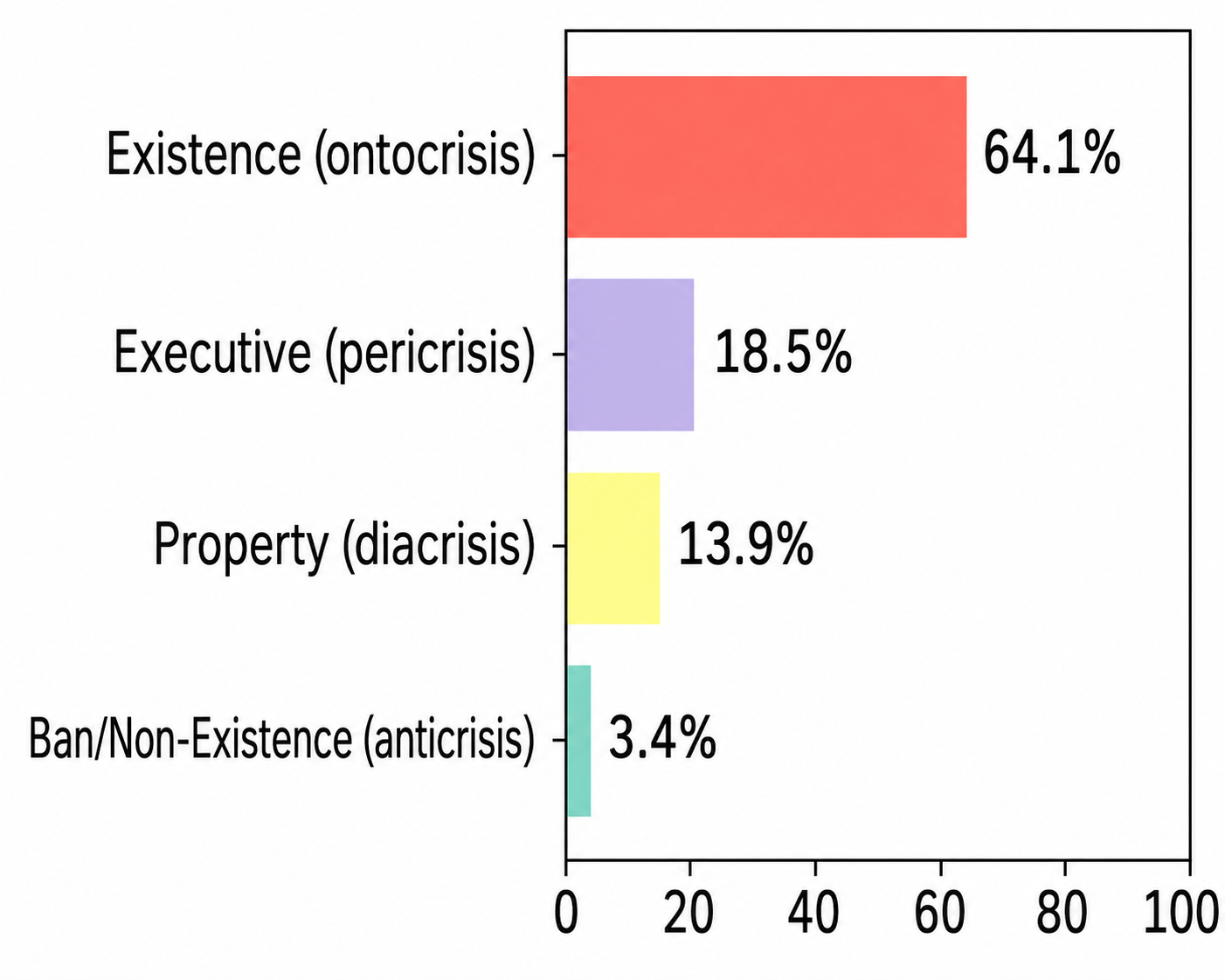}
        \caption{Kruchten's taxonomy.}
        \label{fig:krutchen}
    \end{subfigure}
    \hfill
    \begin{subfigure}[t]{0.32\textwidth}
        \centering
        \includegraphics[width=\linewidth]{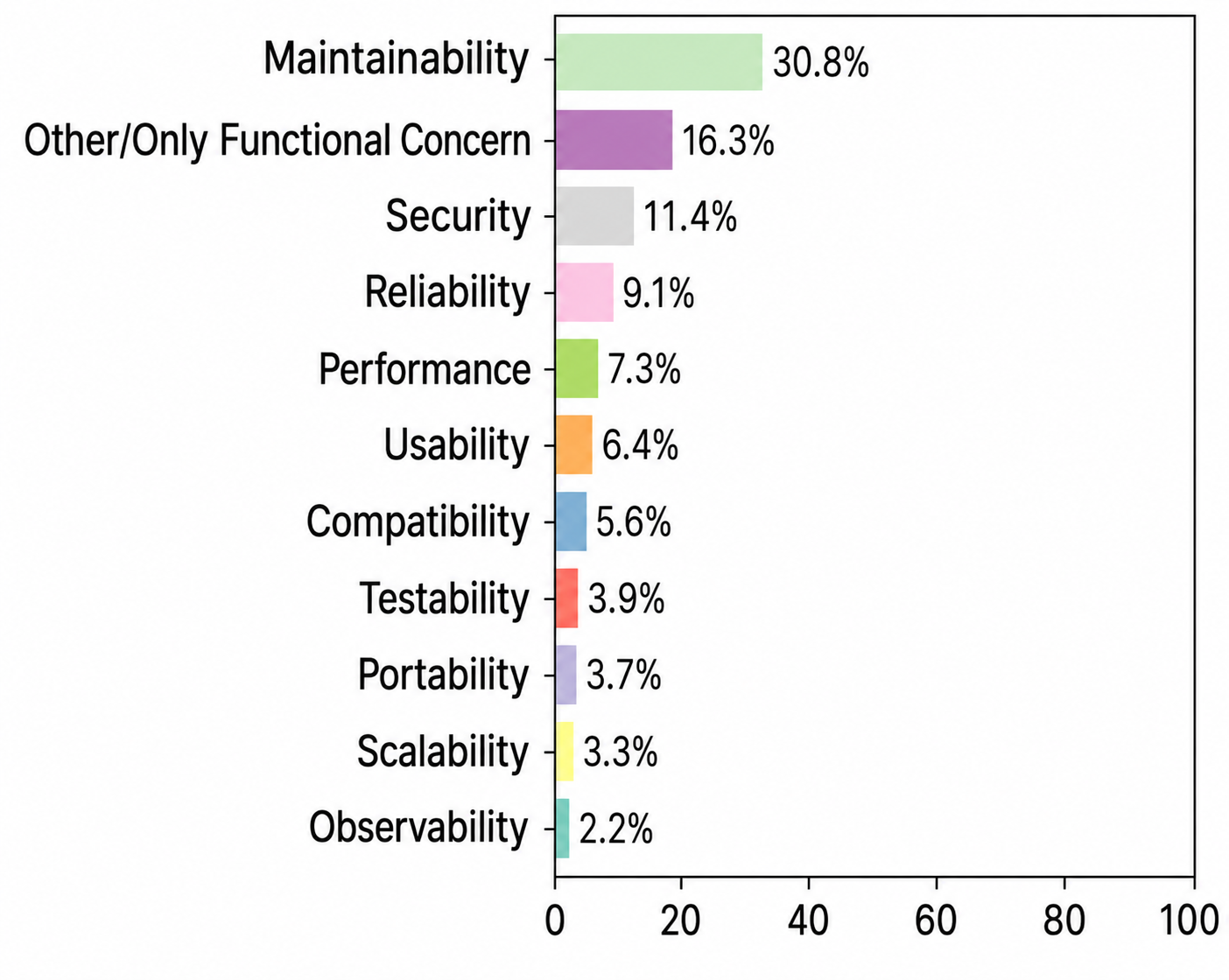}
        \caption{Quality-attribute concerns.}
        \label{fig:qualities}
    \end{subfigure}
    \hfill
    \begin{subfigure}[t]{0.32\textwidth}
        \centering
        \includegraphics[width=\linewidth]{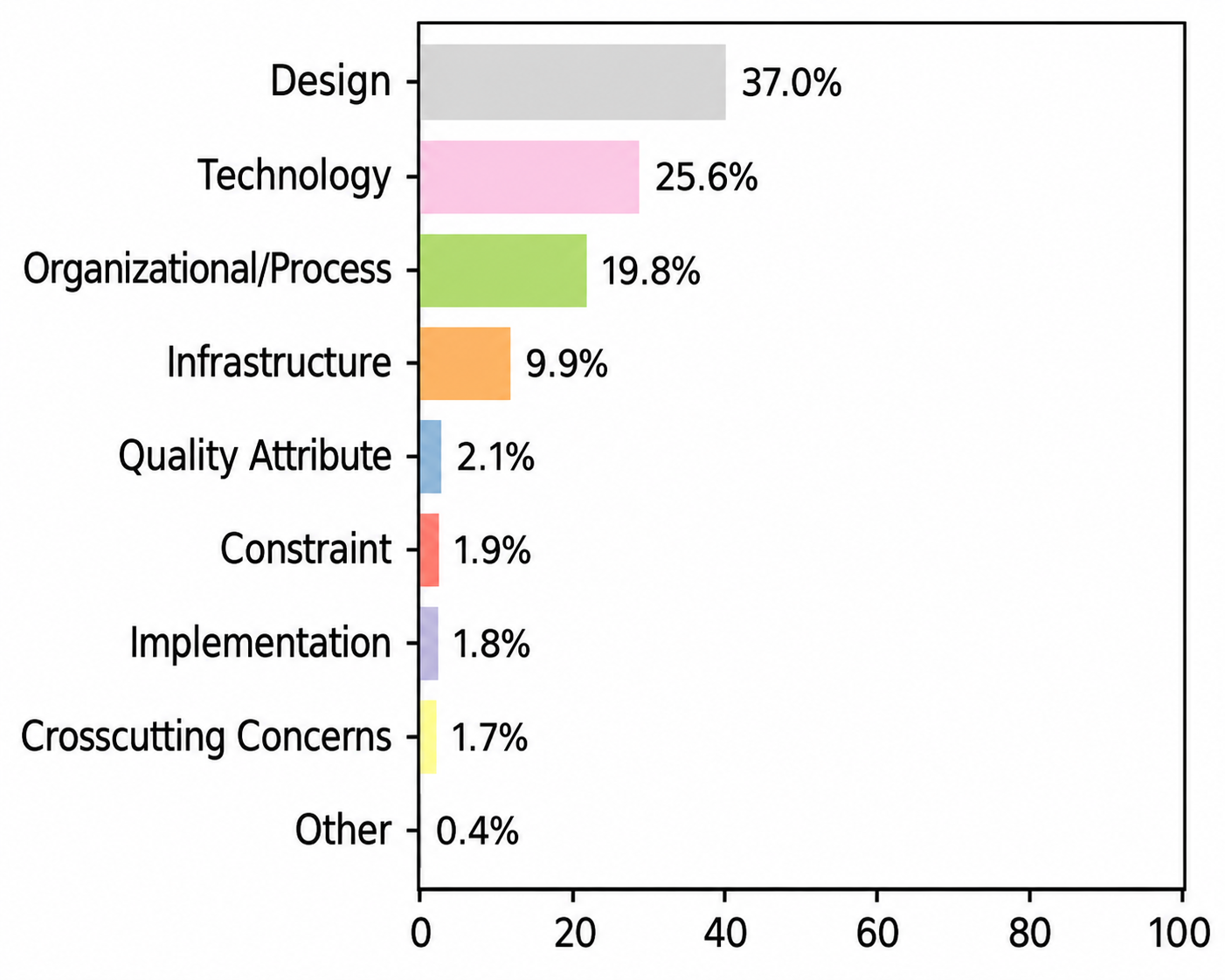}
        \caption{Zimmerman's taxonomy.}
        \label{fig:zimmermann}
    \end{subfigure}\vspace{-0.3cm}
    
    \caption{Overview of ADR classification results across three analysis dimensions.}\vspace{-0.75cm}
    \label{fig:adr-results}
\end{figure*}

We analyzed ADRs according to the three taxonomies introduced in Section~\ref{sec:decision-taxonomy}. Figure~\ref{fig:krutchen} summarizes the results for Kruchten’s taxonomy. When multiple decision types were plausible, we kept the category with the highest LLM confidence. The results show a strong dominance of \textit{existence} decisions ($64\%$), followed by \textit{executive} ($18\%$), \textit{property} ($14\%$), and a small number of \textit{non-existence} decisions.

This distribution is broadly aligned with prior work \cite{miesbauer2013classification}, reinforcing the idea that ADRs primarily document design choices about what should exist in the system. In our dataset, existence decisions were frequently associated with infrastructure, authentication, API management, testing, and repository organization. Property decisions were less frequent and often related to error handling or identifier usage, while executive decisions tended to involve development practices such as CI/CD or documentation processes. Although rare, non-existence decisions were also present, typically documenting technologies, mechanisms, or dependencies that should be avoided. This finding shows evidence that architects do rely on non-existence decisions when necessary.

Despite our prompt improvement efforts, the LLM classifier sometimes confused executive and existence decisions, which can be due to the lack of context (in the ADR itself) about the product/system and the surrounding business/development environment for it. For a better classification, the LLM prompt should incorporate the project context, which is often outside the ADR text.

Figure~\ref{fig:qualities} shows the quality-attribute concerns most frequently captured in ADRs. The most prevalent were \textit{Maintainability}, \textit{Security}, \textit{Reliability}, and \textit{Performance}, in addition to ADRs classified as primarily \textit{Functional}. We observed a gray area to distinguish decisions with a strong functional component and very limited quality emphasis from decisions where quality attributes were more prevalent. Also, we observed that maintainability often absorbed related concerns such as compatibility or portability, while some ADRs occupied a gray area between functional intent and explicit quality emphasis.

Crosslinking Kruchten’s taxonomy with quality attributes revealed that most existence decisions were associated with maintainability, security, reliability, and functionality. In practice, these concerns were often reflected in ADRs about testing, logging, secure communication, authentication, queues, caching, or technology integration. This suggests that ADRs are commonly used to justify not only structural choices, but also operational and quality-related trade-offs.

The results for Zimmermann’s taxonomy are shown in Figure~\ref{fig:zimmermann}. Nearly $40\%$ of the ADRs were classified as \textit{Design} decisions, followed by \textit{Technology} decisions (about one quarter of the corpus). 
With respect to Kruchten's taxonomy, we corroborated that most existence decisions referred to design or technology concerns.
Another notable finding is that approximately $20\%$ of ADRs were classified as \textit{Organizational/Process} decisions, covering concerns such as CI/CD, change logs, approval workflows, testing, and feature flags. This highlights that ADRs are often used to document not only product-level architectural choices, but also decisions related to development and governance practices.

Overall, the taxonomy-based analysis confirms that ADRs are predominantly used to capture existence-oriented, design- and technology-driven decisions, while also serving for documenting quality concerns and organizational processes.

\vspace{-0.5cm}\subsection{RQ\#3 – Usage of ADR Template Sections}\vspace{-0.25cm}

\begin{figure}[t]
\centering
\includegraphics[width=0.6\linewidth]{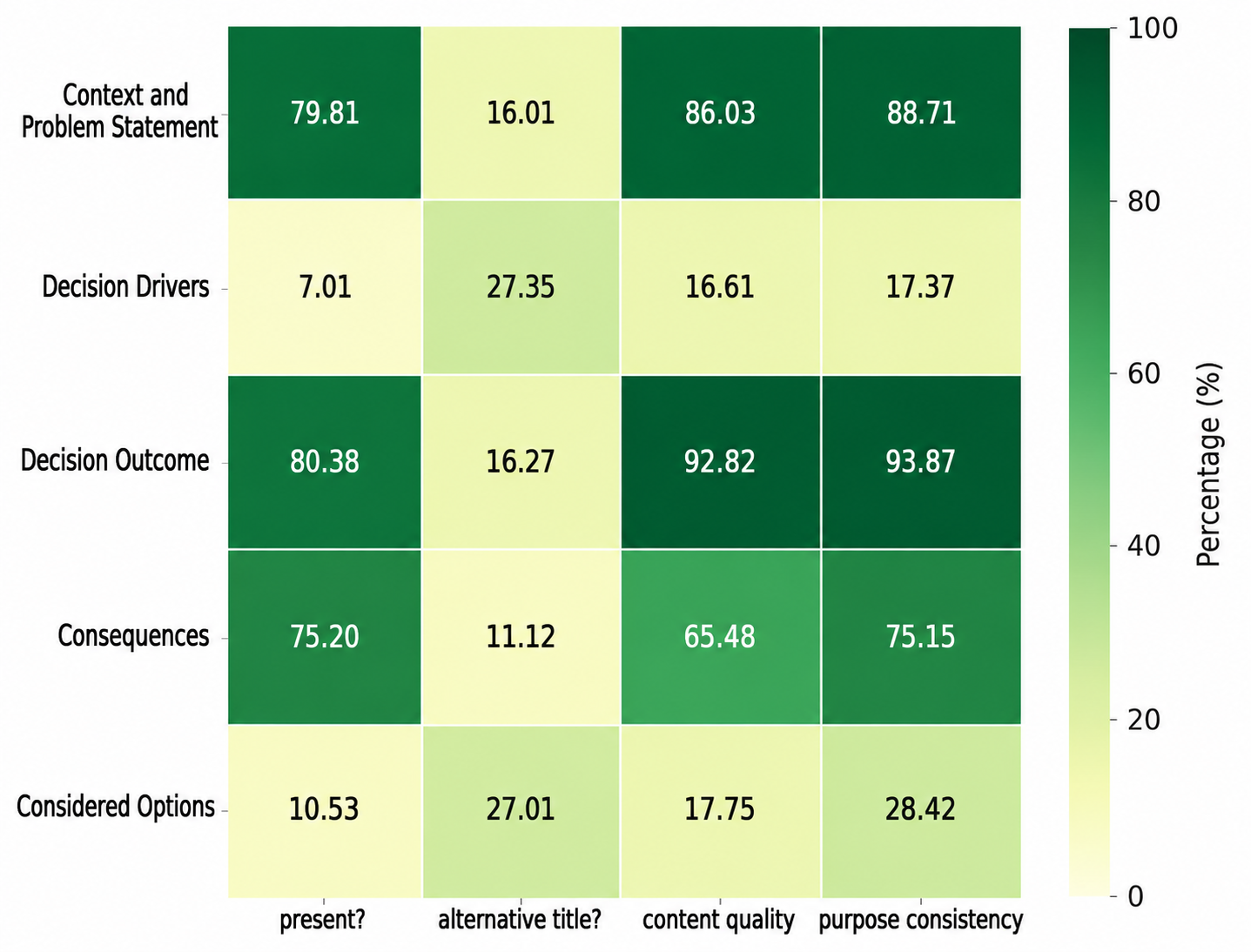}
\vspace{-0.2cm}
\caption{Degree of compliance of ADR sections to 
evaluation criteria (e.g., section presence, content quality, etc.).}\label{fig:compliance}\vspace{-0.85cm}
\end{figure}

To assess the structural compliance of ADRs with the MADR template, we used an LLM-as-a-judge approach to determine whether each section was present and whether its content fulfilled the intended purpose. For each criterion, we computed the percentage of ADRs complying with the condition under analysis. The results are shown in Figure~\ref{fig:compliance}, where higher percentages indicate stronger adherence to the expected template structure. Cases labeled as \textit{Partial} for purpose consistency were conservatively treated as satisfactory.

The \textit{Context and Problem Statement} and \textit{Decision Outcome} sections were generally present, correctly placed, and contained relevant content. The \textit{Consequences} section (for the main decision) also appeared relatively often, although with lower content quality. In contrast, \textit{Decision Drivers} were frequently misplaced (e.g., in the context or decision sections) and often lacked substantive content. The most problematic section was \textit{Considered Options}, which is mandatory in the MADR template (see Fig. \ref{fig:adr-template}). This section was absent or misplaced in most ADRs and, when present, often failed to fulfill its intended purpose.

These results suggest recurring structural weaknesses in ADR authoring. In particular, the low quality and frequent omission of \textit{Decision Drivers} and \textit{Considered Options} point to a limited documentation of rationale and alternatives. This weakens the traceability of architectural reasoning and may contribute to the loss of architectural knowledge over time.

\vspace{-0.55cm}\section{Discussion}\vspace{-0.25cm}

Our findings provide insights on how ADRs are actually documented in terms of common topics, and organized around the MADR template, which uncovers opportunities for enhancing the ADR authoring process. The process of building LLM-based classifiers also sheds light on ambiguity cases when categorizing ADRs. Furthermore, our content-based evaluation highlights which ADR sections and decision types are most often under-documented. 
Overall, ADRs in open-source repositories frequently document technology, infrastructure, and process-related decisions, but often under-document rationale-related elements such as decision drivers and considered alternatives. 
This suggests that ADRs are valuable artifacts, but that their current usage does not always reflect established documentation practices for architectural knowledge. 
This is especially relevant because these omitted elements are precisely the ones that support architectural traceability, onboarding, system evolution, and long-term maintainability.

For practitioners, these findings suggest that ADRs from open-source repositories should not necessarily be treated as exemplary documentation artifacts. Teams adopting ADR practices may unintentionally inherit incomplete documentation habits that emphasize solution choices while neglecting rationale and trade-offs, hindering architectural understanding and making past decisions harder to revisit. ADR authoring may therefore benefit from more systematic review practices (e.g., during pull requests) and stronger attention to omitted sections such as rationale, alternatives, and consequences. Our analysis also challenge a common assumption that ADRs mainly capture ``pure" architectural concerns: in practice, they are often used to document adjacent but relevant aspects such as deployment, implementation, and development processes.
Consequently, existing AK taxonomies and tooling may only partially reflect the actual documentation practices followed by development teams.

The frequent omission of \textit{Decision Drivers} and \textit{Considered Options} may reflect practical constraints such as time pressure, lightweight documentation habits, or the perception that rationale is already captured elsewhere (e.g., code reviews or issue trackers). Although these explanations remain hypothetical, they point to research opportunities for qualitative studies aimed at understanding how practitioners perceive the costs and benefits of ADR documentation.

For researchers and tool builders, our findings highlight opportunities to improve ADR authoring support. For example, ADR tooling could suggest missing sections, surface likely quality concerns, or automatically attach metadata inferred from decision taxonomies and topic models. Such metadata could also help maintain and query a project’s architectural knowledge base over time.  More broadly, our results motivate longitudinal and industrial studies to better understand how ADR practices evolve across project maturity levels, organizational contexts, and tooling ecosystems.

\vspace{-0.55cm}\subsection{Threats to Validity}\vspace{-0.25cm}

As in any empirical study, several factors may affect the validity of our findings. Following \cite{vidoni2022systematic}, we summarize the main threats below.

\vspace{-0.35cm}\paragraph*{Internal Validity}
Our detection of ADR structure relied primarily on section headings, which may miss custom or informal formats, while decision-type and section-consistency classification depended on LLM prompting. Although this enabled scalable analysis, outputs may still be affected by prompt design, ambiguous phrasing, or project-specific terminology. Two authors independently reviewed subsets of the LLM outputs and found the resulting labels generally consistent with the intended interpretations.

\vspace{-0.35cm}\paragraph*{Construct validity}
To ensure that we were measuring the intended constructs, we grounded our analyses in established decision taxonomies and predefined evaluation criteria. Although ADRs may simultaneously involve multiple concerns or quality attributes, we operationalized classification primarily as a single-label task by assigning the dominant category to each ADR. 
This choice enabled quantitative comparison and aggregation at scale, while acknowledging that it may abstract away secondary categories for some decisions (e.g., typically quality attributes). Nonetheless, the LLM structured response considered slots for alternative decision categories, which will be integrated into future work.

\vspace{-0.35cm}\paragraph*{External Validity}
The dataset is limited to English-language ADRs from open-source GitHub repositories. Thus, findings may not generalize to non-English, industrial, or private software projects, where documentation practices may differ. The scale and diversity of the dataset provide a meaningful empirical basis for understanding ADR usage in open-source settings. Nevertheless, given that $40\%$ of the projects are from 2020 (the last year of the dataset), this rapid increase in ADR usage \cite{buchgeher2023using} calls for additional studies from 2021 to the present.
Our analysis focused on ADRs written in Markdown format. As a result, ADRs documented in alternative formats (e.g., wikis, issue trackers) were not included, which may limit the generalizability of the findings to other documentation ecosystems.

\vspace{-0.35cm}\section{Conclusions and Future Work}\vspace{-0.3cm}

This paper proposed an automated approach for mining and analyzing ADRs, which we applied to a large dataset of open-source repositories. Our study provides empirical evidence about what ADRs actually document in practice. In particular, we found that ADRs are dominated by existence-oriented, design- and technology-driven decisions, while also frequently capturing infrastructure, deployment, and process-related concerns. We also observed recurring weaknesses in the use of the MADR template, especially in sections related to rationale and alternatives. 
By answering ``\textit{what}" is discussed in ADRs, this research seeks to improve AK management by identifying misalignments between the expected and actual contents of ADRs. 
Certainly, determining the best ADR label for a given taxonomy is debatable for certain categories, even for human users. While still fallible, an LLM-approach seems to work better than embeddings of ADR texts, because the LLM can reason about the ADR semantics, if provided with adequate prompting and context. Furthermore, we contribute an automated processing pipeline and methodology, which can be tailored to other architectural tasks (e.g., architecture conformance) or domain-specific needs. 

Our findings open up several directions for future work.  
First, qualitative studies with practitioners could help explain \textit{why} certain ADR documentation patterns emerge. Second, extending the analysis to industrial or private repositories would improve external validity. Third, ADRs could be enriched with inferred metadata or linked into semantic graphs to better support architectural knowledge management. Finally, LLM-powered tools could help practitioners write (better) ADRs from basic inputs and leverage project context \cite{diaz2024helping, dhar2025draft}.

\vspace{-0.45cm}\subsubsection{Data Availability} All data and source code used in this study are available at \url{https://github.com/tommantonela/ADRminer}.
 
\vspace{-0.45cm}\subsubsection{Acknowledgments.} This work was partially supported by PICT-2021-00757 project (Argentina).
\vspace{-0.25cm}\renewcommand\bibsection{\section*{\refname}}
\bibliographystyle{splncs04}
\bibliography{references}

\end{document}